\documentclass[a4paper]{appolb}
\usepackage{epsfig}
\usepackage[T1]{fontenc}

\usepackage{amsmath}
\usepackage{amssymb,mathrsfs}
\usepackage{floatflt}
\usepackage{wrapfig}
\def\slashchar#1{\setbox0=\hbox{$#1$}
   \dimen0=\wd0 \setbox1=\hbox{/} \dimen1=\wd1
   \ifdim\dimen0>\dimen1 \rlap{\hbox to \dimen0{\hfil/\hfil}} #1
   \else  \rlap{\hbox to \dimen1{\hfil$#1$\hfil}} / \fi}

\begin{document}
\title{Coherent Neutrino Scattering%
\thanks{Presented by M. Valverde at the 45th Winter School in Theoretical Physics ``Neutrino Interactions: from Theory to Monte Carlo Simulations'', L\k{a}dek-Zdr\'oj, Poland, February 2--11, 2009.}%
}
\author{M. Valverde
\address{Research Center for Nuclear Physics (RCNP), Osaka University, Ibaraki~567-0047, Japan}
\and
 J.~E.~Amaro
\address{Dpto. F\'\i sica At\'omica, Molecular y Nuclear, Univ. de Granada,\\ E-18071~Granada, Spain}
\and
E. Hernandez
\address{Grupo de F\'\i sica Nuclear, Dpto.
de F\'\i sica Fundamental e IUFFyM,\\ Facultad de Ciencias, E-37008
Salamanca, Spain}
\and
J. Nieves
\address{Instituto de F\'\i sica Corpuscular (IFIC), Centro Mixto
  CSIC-Univ. de Valencia, Institutos de Investigaci\'on de
  Paterna, Aptd. 22085, E-46071 Valencia, Spain}
}
\maketitle
\begin{abstract}
We present a microscopic model for coherent pion production off nuclei
induced by neutrinos. This model is built upon a model for single
nucleon processes that goes beyond the usual $\Delta$ dominance by
including non resonant background contributions. We include nuclear
medium effects: medium corrections to $\Delta$ properties and outgoing
pion absortion via an optical potential. This results in major
modifications to cross sections for low energy experiments when
compared with phenomenological models like Rein--Sehgal's.
\end{abstract}


A proper understanding of neutrino-induced pion production off nuclei
is very important in the analysis of neutrino oscillation
experiments. For instance, $\pi^0$ production by neutral currents (NC)
is the most important $\nu_\mu$-induced background to
$\nu_\mu\to\nu_e$ oscillation experiments, \cite{AguilarArevalo:2008xs}. Similarly, $\pi^+$ production by charged
currents (CC) is an important source of background in $\nu_\mu\to
\nu_x$ disappearance searches~\cite{Hiraide:2008eu}.  We will follow \cite{Amaro:2008hd} to
describe the coherent CC pion production reaction induced by neutrinos
\begin{equation}
  \nu_l (k) +\, A_Z|_{gs}(p_A)  \to l^- (k^\prime) +
  A_Z|_{gs}(p^\prime_A) +\, \pi^+(k_\pi)  \, ,
\label{eq:reac}
\end{equation}
where the nucleus is left in its ground state, in contrast to
incoherent reactions where the nucleus is broken or left on an excited
state.

We build upon a microscopic model for the single nucleon process ($\nu
N\to l^-N\pi^+$). We sum coherently the contribution of all nucleons
on the initial nuclei, which is modeled after a Fermi gas in Local
Density Approximation.
Coherent $\pi$ production is most sensitive to the Fourier transform of
the nuclear density for momentum $\vec{q}-\vec{k}_\pi$, which gets its
maximum value when $\vec{q}$ and $\vec{k}_\pi$ are parallel. For this
particular kinematics the vector contribution to the single nucleon
($W+N\to N\pi$) currents, which is purely transverse $\vec{k}_\pi
\times \vec{q}$, vanishes unlike the axial contribution.  This
dominance of the axial contributions is exploited through the PCAC
hypothesis by the Rein--Sehgal (RS)
model~\cite{Rein,Hernandez:2009vm}, to relate the
neutrino coherent pion production cross section with the pion-nucleus
elastic differential one.



For the elementary process we use the model derived in
\begin{wrapfigure}{r}{0.45\textwidth}
\centering
\includegraphics[width=0.45\textwidth]{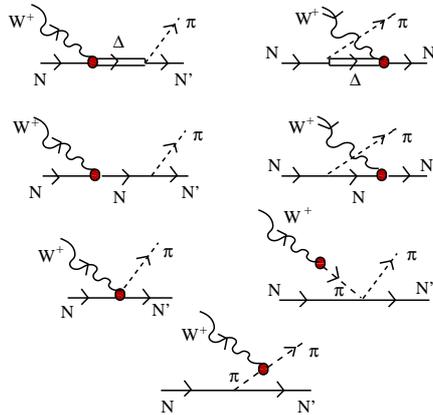}
\caption{\footnotesize Model for the $W^+N\to N^\prime\pi$ reaction.
  The circle in the diagrams stands for the weak vertex.}
\label{fig:diagramas1}
\end{wrapfigure}
Ref.~\cite{Hernandez:2007qq}, see Fig.~\ref{fig:diagramas1}.  In
addition to the $\Delta(1232)$ pole ($\Delta P$) (first row) mechanism
the model includes background terms required by chiral symmetry:
nucleon (second row) pole terms ($NP$, $CNP$) contact ($CT$) and pion
pole ($PP$) contribution (third row) and pion-in-flight ($PF$) term.
Background terms turn out to be very important and because of them,
the flux-averaged $\nu_\mu p\to \mu^- p \pi^+$ ANL cross section
~\cite{Barish:1978pj,Radecky:1981fn} is described with an axial form
factor where the dominant $C_5^A$ nucleon-to-$\Delta$ axial form
factor was fit to data resulting in $C_5^A(0)=0.867$ and $M_{A\Delta}=
0.985$ GeV. This value for $C_5^A(0)$ is significantly smaller than
the value of about $1.2$ deduced from the Golberger-Treiman relation
(GTR) used in PCAC-based approaches like RS.

When applied to a coherent process in finite nuclei we find
that the $NP$ and $CNP$ nucleon pole term contributions partially
cancel each other,
that the $PF$ term does not contribute to the coherent cross section
\begin{wrapfigure}{r}{0.5\textwidth}
\centering
\includegraphics[width=0.5\textwidth]{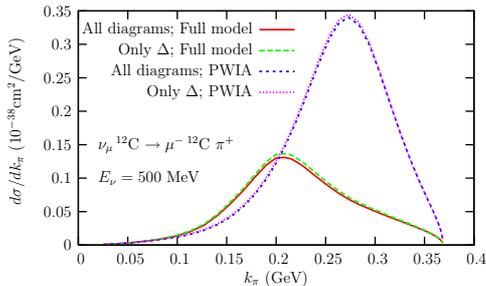}
\caption{\footnotesize Pion momentum differential LAB
  cross section, with and without background terms.}
\label{fig:diagramas2}
\end{wrapfigure}
and the $CT$ and $PP$ terms vanish for isospin
symmetric nuclei. As seen in Fig.~\ref{fig:diagramas2} the effect of
the background terms, both in
the plane wave impulse approximation (PWIA) and in the full model calculation, is very small.  Thus, we
predict cross sections around a factor of $(1.2/0.9)^2\sim 2$ smaller
than approaches assuming GTR. In the following we will always use the
full model of Ref.~\cite{Hernandez:2007qq} with $C_5^A(0)=0.867$ and
$M_{A\Delta}= 0.985$ GeV.

\begin{figure}[htb]
\begin{center}
\makebox[0pt]{\includegraphics[width=0.5\textwidth]{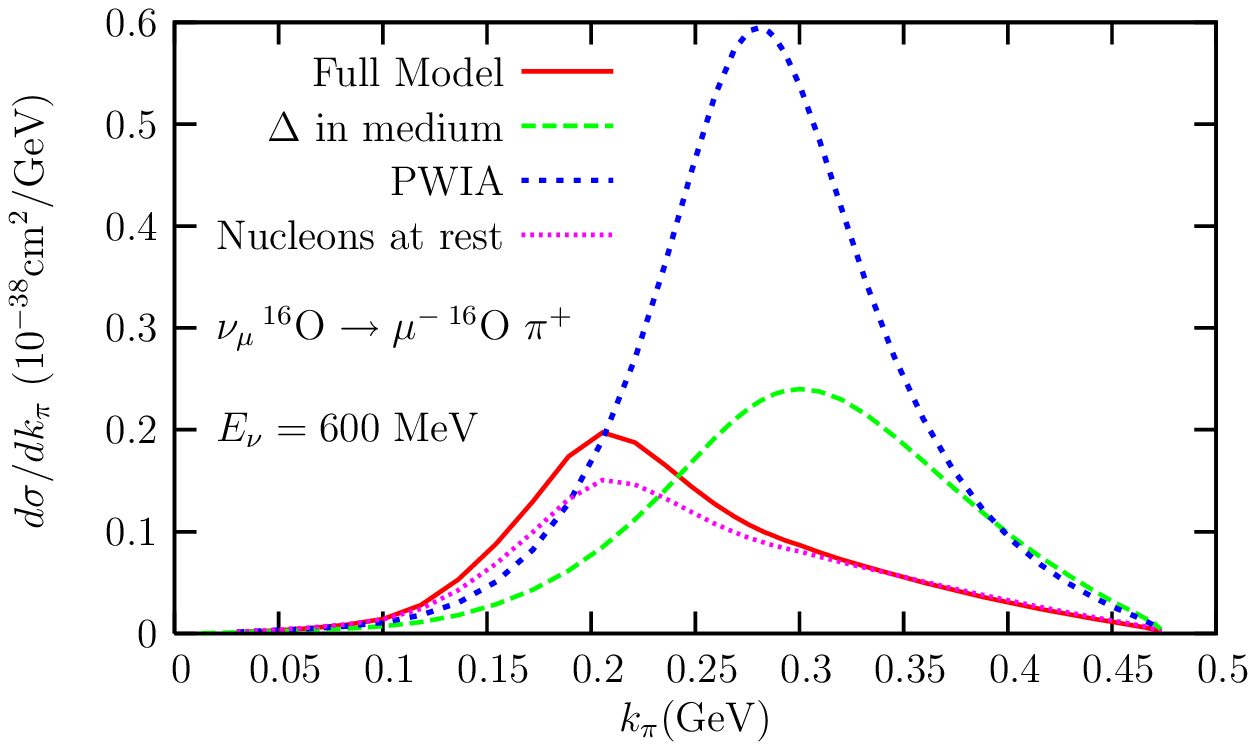}
              \includegraphics[width=0.5\textwidth]{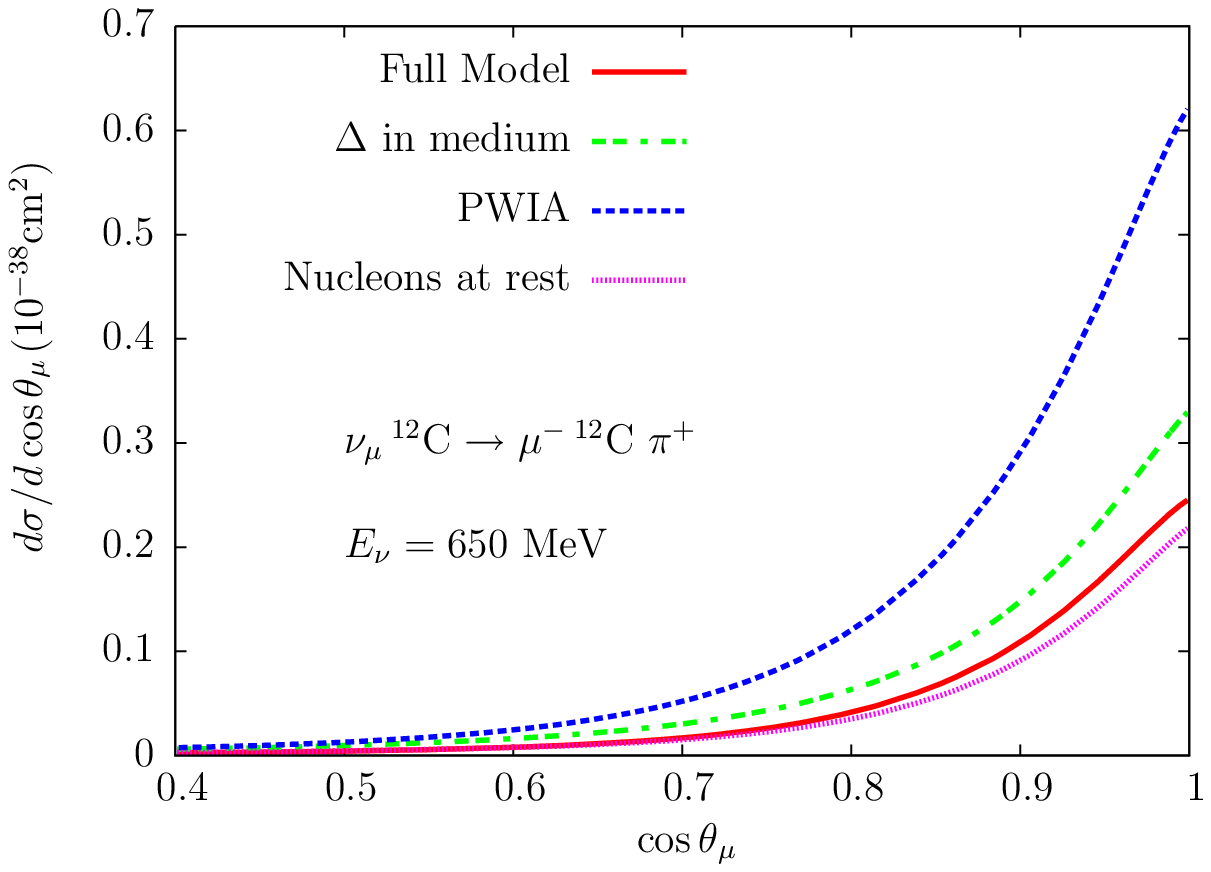}}
\end{center}
\caption{\footnotesize Right panel: Pion momentum differential cross section in the
  LAB frame.
Left panel: Pion angular differential cross section.
}\label{fig:dspion}
\end{figure}

Nuclear medium corrections to the dominant $\Delta$ diagram are
considered by including the self-energy of the $\Delta$ in the medium,
Ref.~\cite{Oset:1987re}.  Another major nuclear medium effect is pion
distortion, which is taken this into account by replacing the plane
wave 
with a pion wave function
incoming solution of a Klein-Gordon equation with a microscopic optical
potential, Ref.~\cite{Nieves:1991ye}.  In left panel of
Fig.~\ref{fig:dspion} we show the pion momentum distribution (LAB) for
CC coherent pion production, in the peak energy region of the T2K
experiment. Including $\Delta$ in-medium self-energy (long-dashed
line) reduces the PWIA results (short-dashed line).  Further inclusion
of pion distortion (full model, solid line) reduces the cross section,
and the peak is shifted towards lower energies.
The total cross section reduction is around $60\%$. Medium and pion
distortion effects in coherent pion production were already evaluated
in Refs.~\cite{AlvarezRuso}. However, the authors of these references
neglected the nucleon momenta in the Dirac spinors. The effect of this
approximation (nucleons at rest, dotted line) results in a $\sim 15$\%
decrease of the total cross section.  In the right panel of
Fig.~\ref{fig:dspion} we show the pion angular LAB distribution with
respect to the incoming neutrino direction. The reaction is very
forward peaked, as expected due to the nucleus form factor.  The
angular distribution profile keeps its forward peaked behavior after
introduction of nuclear medium effects.

\begin{figure}[htb]
\begin{center}
\makebox[0pt]{\includegraphics[width=0.5\textwidth]{epicosteta.eps}
              \includegraphics[width=0.5\textwidth]{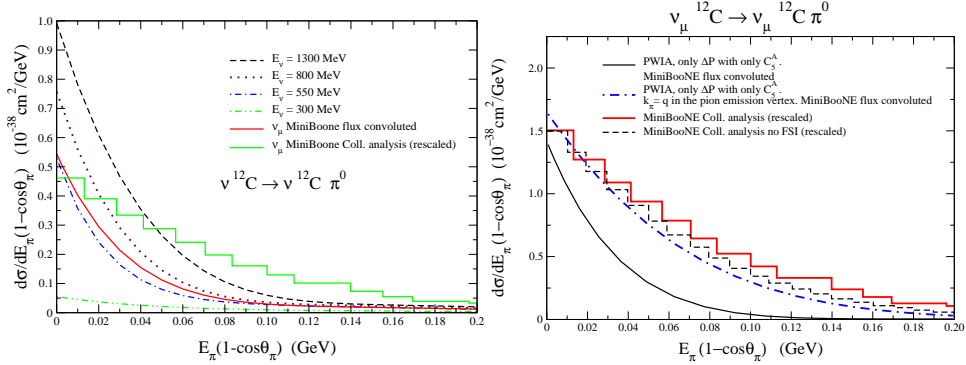}}
\end{center}
\caption{\footnotesize Laboratory $E_\pi (1-\cos\theta_\pi)$, at
  MiniBooNE energies. In the left panel we use our full model including
  full nuclear corrections. In the right panel, we show results from
  the $C_5^A$ axial contribution of the $\Delta P$ mechanism,
  neglecting pion distortion and $\Delta$ in medium effects.  We
  display the MiniBooNE published histogram (solid), conveniently
  scaled down, from Ref.~\protect\cite{AguilarArevalo:2008xs}
  and MiniBooNE results (dashed histogram) obtained by turning off the
  NUANCE FSI of the outgoing pion (G. Zeller, private communication).
}\label{fig:fig_dxnc}
\end{figure}

We examine in Fig.~\ref{fig:fig_dxnc} the NC differential cross
section with respect to the variable $E_\pi (1-\cos\theta_\pi)$,
proposed by MiniBooNE.  Our prediction is appreciably narrower than
that displayed in Fig.~3b of Ref.~\cite{AguilarArevalo:2008xs}. The
MiniBooNE analysis relies on the RS model, so we try to understand the
differences between this and our model. RS's expression for the
coherent $\pi^0$ production cross section was deduced in the parallel
configuration, for which the $k_\mu$ and $k^\prime_\mu$ four momenta
are proportional ($q^2=0$) and $\vec{k}_\pi \approx \vec{q}$ is
assumed everywhere except in the nuclear form factor. Thus, the RS
differential cross section depends on $\cos\theta_\pi$ or $t$ only
through the nuclear form factor and any further $\cos\theta_\pi$ or
$t$ behaviour induced by the dependence of the amplitudes on $k_\pi$
is totally neglected.  This is a good approximation at neutrino
energies above $2$~GeV. However, at the energies relevant in the
MiniBooNE and T2K experiments non parallel configurations become
important, and the RS model less reliable.
We have re-derived RS's expression within our model by considering
only the dominant axial part of the $\Delta P$ process ($\sim C_5^A$),
neglecting nuclear medium corrections and replacing $k_\pi$ by $q$ in
the pion emission vertex.  In the right panel of
Fig.~\ref{fig:fig_dxnc} we see that the new $E_\pi(1-\cos\theta_\pi)$
distribution is significantly wider than that obtained without
implementing this replacement and that it reasonably describes the
MiniBooNE published distribution (solid histogram). The
agreement is much better when compared with some preliminary MiniBooNE
results (dashed histogram) obtained with a different treatment of the
outgoing pion distortion. This calculation shows the uncertainties
associated to the $t=0$ approximation at low energies.

Pion distortion induces some additional discrepancies.  MiniBooNE
implement this effects through a Monte Carlo cascade model for the
$\pi$ propagation in medium.  However, coherent cross sections cannot
be calculated from a Monte Carlo cascade algorithm, because the
coherent production is a one step process and by using a Monte Carlo
algorithm we break the coherence of the process.  Nevertheless, one
could still reasonably estimate the total coherent cross section from
the NUANCE FSI cascade if it is used to eliminate from the flux not
only those pions which get absorbed or suffer inelastic processes but
also those that undergo QE steps. To our knowledge, these latter
events are accounted for in the MiniBooNE analysis, despite not being
coherent.  In our calculation the imaginary part of the pion-nucleus
potential removes from the flux of the outgoing pions those that are
absorbed or undergo QE interactions.  We try to estimate this effect
by switching off the QE contribution to the pion-nucleus optical
potential induced by elastic pion-nucleon collisions, and using an
optical potential with an imaginary part due to absorption and
inelastic channels alone.  For the MiniBooNE flux averaged cross
section we find a $20$\% enhancement (see NC* entry in
Table~\ref{tab:res}) in good agreement with the effects observed by
turning off the NUANCE FSI.  We conclude that the RS model is not as
reliable for MiniBooNE and T2K experiments as for $\nu$ energies above
$2$~GeV. Our model provides an $E_\pi (1-\cos\theta_\pi)$ distribution
much more peaked, and thus it might improve the description of the
first bin value in Fig.~3b of
Ref.~\cite{AguilarArevalo:2008xs}. Moreover, the drastic change in the
$E_\pi (1-\cos\theta_\pi)$ distribution shape might produce some
mismatch between the absolute normalization of the background,
coherent and incoherent yields in the MiniBooNE analysis.

\begin{table}[htb]
\begin{center}
    \begin{tabular}{lcccc}\hline
Reaction     & Exp. & $\bar \sigma$[$10^{-40}$cm$^2$] & $\sigma_{\rm exp}$[$10^{-40}$cm$^2$] &$E_{\text{max}}$ [MeV] \\\hline
CC\phantom{*} $\nu_\mu + ^{12}$C    & K2K        &   $4.68$    & $<7.7 $~\cite{Hasegawa:2005td} & $1.80$\\
CC\phantom{*} $\nu_\mu + ^{12}$C    & MiniBooNE  &   $2.99$    &    & $1.45$ \\
CC\phantom{*} $\nu_\mu + ^{12}$C    & T2K        &   $2.57$    &    & $1.45$ \\
CC\phantom{*} $\nu_\mu + ^{16}$O    & T2K        &   $3.03$    &    & $1.45$ \\
NC\phantom{*} $\nu_{{\mu}} + ^{12}$C    & MiniBooNE  & $1.97$  & $7.7\pm1.6\pm3.6$~\cite{Raaf:2005up} & $1.34$ \\
NC* $\nu_{{\mu}} + ^{12}$C              & MiniBooNE  & $2.38$  & $7.7\pm1.6\pm3.6$~\cite{Raaf:2005up} & $1.34$ \\
NC\phantom{*} $\nu_{{\mu}} + ^{12}$C    & T2K        & $1.82$  &    & $1.34$ \\
NC\phantom{*} $\nu_{{\mu}} + ^{16}$O    & T2K        & $2.27$  &    & $1.35$ \\
CC\phantom{*} $\bar\nu_\mu + ^{12}$C    & T2K        & $2.12$  &    & $1.45$ \\
NC\phantom{*} $\bar\nu_{{\mu}} + ^{12}$C& T2K        & $1.50$  &    & $1.34$ \\ \hline
    \end{tabular}
\end{center} 
  \caption{\footnotesize Coherent pion production total cross
    sections. 
}
\label{tab:res} 
\end{table}
In Table~\ref{tab:res} we show our predictions for the MiniBooNE, K2K
and T2K~\cite{Kato:2007zzc} flux averaged cross sections. Since our
model neglects all resonances above the $\Delta$, our predictions
become less reliable when the energy increases, so we set up a maximum
neutrino energy in the flux convolution $E_{\text {max}}$, neglecting
the long tail of the $\nu$ flux. Up to these energies, one can assume
$\Delta$ dominance and still cover about $90$\% of the total flux
($65$\% for T2K antineutrino flux). We expect corrections (higher
cross sections) of around $20$--$30$\% to our results for MiniBooNE
and T2K (larger for K2K). Our prediction lies well below the K2K upper
bound, while being notably smaller than that given in
\cite{Raaf:2005up} for NC MiniBooNE. However, notice the previous
discussion on RS model, which is being used in the MiniBoone analysis.
The K2K cross section and the value quoted in Ref.~\cite{Raaf:2005up}
seems somehow incompatible with the approximate relation $\sigma_{\rm
  CC} \approx 2 \sigma_{\rm NC}$, expected from $\Delta-$dominance and
neglecting finite muon mass effects.  
%

%

\end{document}